\documentclass[11pt]{article}
\usepackage{amssymb}
\usepackage{epsfig}
\usepackage{cite}
\usepackage{rotating}

\newcommand{\epem}{\ensuremath{\mathrm{e}^+\mathrm{e}^-}}

\newcommand{\Zz}{\ensuremath{{\mathrm{Z}^0}}}

\newcommand{\dedx}{\ensuremath{\mathrm{d}E/\mathrm{d}x}}

\newcommand{\definmath}[2] {\def#1{\ifmmode#2\else$#2$\fi}}
\definmath{\degree}  {^\circ}

\definmath{\Pq}      {\mathrm{q}}
\definmath{\Paq}  {\overline{\mathrm{q}}}
\definmath{\Pu}      {\mathrm{u}}
\definmath{\Pau}  {\overline{\mathrm{u}}}
\definmath{\Pd}      {\mathrm{d}}
\definmath{\Pad}  {\overline{\mathrm{d}}}
\definmath{\Ps}      {\mathrm{s}}
\definmath{\Pas}  {\overline{\mathrm{s}}}
\definmath{\Pc}      {\mathrm{c}}
\definmath{\Pac}  {\overline{\mathrm{c}}}
\definmath{\Pb}      {\mathrm{b}}
\definmath{\Pab}  {\overline{\mathrm{b}}}
\definmath{\Pt}      {\mathrm{t}}
\definmath{\Pat}  {\overline{\mathrm{t}}}

\definmath{\Pgtp} {\tau^{+}}        
\definmath{\Pgtm} {\tau^{-}}        
\definmath{\Pgtpm}   {\tau^{\pm}}         

\def\etal{\mbox{{\it et al.}}}

\def\gappeq{\ensuremath{\mathrel{ \rlap{\raise.5ex\hbox{>}}
                      {\lower.5ex\hbox{\sim}}}}}
\def\lappeq{\ensuremath{\mathrel{ \rlap{\raise.5ex\hbox{<}}
                      {\lower.5ex\hbox{\sim}}}}}

\newcommand{\majorsec} {\it High Sector}
\newcommand{\minorsec} {\it Second Sector}
\begin{document}
\begin{titlepage}
%
%
\begin{center}
     \Large
     EUROPEAN ORGANIZATION FOR NUCLEAR RESEARCH 
\end{center}
\begin{flushright}
    \Large
    CERN-PH-EP-2007-019 \\
     12th June 2007\\
\end{flushright}
\bigskip
\begin{center}
    \huge\bf\boldmath
 Search for Dirac Magnetic Monopoles \\ \vspace*{3mm} 
 in \epem\ Collisions  \\ \vspace*{3mm}
 with the OPAL Detector at LEP2
\end{center}
\bigskip
\bigskip

\begin{center}
\LARGE
The OPAL Collaboration \\
\bigskip
\bigskip
\end{center}


\bigskip

\begin{abstract}

This letter describes a direct search for pair produced magnetic monopoles
in \epem\ collisions.
The analysis is based on 62.7~pb$^{-1}$ of data collected with the
OPAL detector 
at an average centre-of-mass energy of $\sqrt{s}$= 206.3~GeV.
The monopole signal was assumed to be characterized by two
back-to-back particles with an
anomalously high ionization energy loss $\dedx$ in the tracking
chambers.
No evidence for production of monopoles was observed. 
Upper limits were obtained on the magnetic monopole pair-production
cross-section ($\sigma$) in the mass range
45~GeV/c$^{2}<m_{M}<102$~GeV/c$^{2}$. The
average limit is $\sigma<0.05$~pb and is essentially independent of 
the magnetic monopole mass. The cross-section limit is derived
at
the 95\% confidence level and is valid for spin-1/2 magnetic
monopoles.
\end{abstract}

\bigskip

\begin{center}
{\large (Submitted to Physics Letters B)}\\
\bigskip
\bigskip
\bigskip
\end{center}
 
\smallskip


\bigskip
\bigskip



 
\end{titlepage}

\begin{center}{\Large        The OPAL Collaboration
}\end{center}\bigskip
\begin{center}{
G.\thinspace Abbiendi$^{  2}$,
C.\thinspace Ainsley$^{  5}$,
P.F.\thinspace {\AA}kesson$^{  7}$,
G.\thinspace Alexander$^{ 21}$,
G.\thinspace Anagnostou$^{  1}$,
K.J.\thinspace Anderson$^{  8}$,
S.\thinspace Asai$^{ 22}$,
D.\thinspace Axen$^{ 26}$,
I.\thinspace Bailey$^{ 25}$,
E.\thinspace Barberio$^{  7,   p}$
% ...
}\end{center}\bigskip
\bigskip
$^{  1}$School of Physics and Astronomy, University of Birmingham,
Birmingham B15 2TT, UK
\newline
$^{  2}$Dipartimento di Fisica dell' Universit\`a di Bologna and INFN,
I-40126 Bologna, Italy
\newline
% ...
$^{  5}$Cavendish Laboratory, Cambridge CB3 0HE, UK
\newline
% ...
$^{  7}$CERN, European Organisation for Nuclear Research,
CH-1211 Geneva 23, Switzerland
\newline
$^{  8}$Enrico Fermi Institute and Department of Physics,
University of Chicago, Chicago IL 60637, USA
\newline
% ...
$^{ 21}$Department of Physics and Astronomy, Tel Aviv University,
Tel Aviv 69978, Israel
\newline
$^{ 22}$International Centre for Elementary Particle Physics and
Department of Physics, University of Tokyo, Tokyo 113-0033, and
Kobe University, Kobe 657-8501, Japan
\newline
% ...
$^{ 25}$University of Victoria, Department of Physics, P O Box 3055,
Victoria BC V8W 3P6, Canada
\newline
$^{ 26}$University of British Columbia, Department of Physics,
Vancouver BC V6T 1Z1, Canada
\newline
% ...
$^{  m}$ current address Bergische Universit\"at, Wuppertal, Germany
\newline
% ...
$^{  p}$ now at The University of Melbourne, Victoria, Australia
\newline
% ...
$^{  s}$ now at University of Antwerpen, Physics Department,B-2610 Antwerpen, 
Belgium; supported by Interuniversity Attraction Poles Programme -- Belgian
Science Policy
\newline
% ...

\bigskip

\section{Introduction}
In 1931 Dirac linked the existence of magnetic monopoles (MMs) with
the quantization of electric charge and postulated the relation
between the elementary
electric charge $e$ of the electron and a basic magnetic charge $g$
\cite{dirac}: 
\begin{equation}
	\begin{array}{ccc}  g = \frac{n\hbar c}{2e} = n g_{D}, & & n =
1, 2, ... \\
	\end{array}
	\label{Eq.Dirac}
\end{equation}
where $n$ is an unknown integer and $g_{D} = \hbar c/2e = 68.5e$ is the
unit Dirac magnetic charge (in the cgs system). 
If free quarks exist, Eq.~\ref{Eq.Dirac} should be modified by
replacing $e$
with $e/3$, which effectively increases $g$ by a factor of 3.
There was no prediction for the monopole mass. A rough estimate,
obtained assuming that the classical monopole radius is equal to the
classical electron radius, yields $m_{M}\approx
g^{2}m_{e}/e^{2}\approx n^{2} \cdot 4700 m_{e}\approx n^{2} \cdot
2.4$~GeV/c$^{2}$.
Since 1931, experimental searches for ``classical Dirac'' monopoles
have been performed at nearly every new high-energy accelerator,
employing a variety of direct and indirect methods \cite{gglp}. By a
classical (Dirac) monopole, we mean a particle without electric
charge or hadronic interactions and with magnetic charge $g$
satisfying the Dirac quantization condition (Eq.~\ref{Eq.Dirac}). 

Within the framework of Grand Unified Theories (GUT) of the
strong and electroweak interactions,
supermassive magnetic monopoles 
with masses $m \geq 10^{16} \,$GeV/c$^{2}$ could have been
produced in the early Universe as intrinsically stable topological
defects at a high energy phase transition that leaves an unbroken
U(1) group \cite{GUTmm}.
At the present time, such monopoles could exist in the penetrating
cosmic radiation as
``fossil'' remnants of that transition.
The detection of such particles would be one of the most spectacular
confirmations of GUT
predictions. 
The most stringent upper limits on an isotropic flux of GUT magnetic
monopoles, assuming monopole masses $m_{M}>10^{16}$~GeV/c$^{2}$, have
been set by the MACRO experiment \cite{MACRO}.  
In some Grand Unified theories values of the monopole mass as low as
$10^{4}$~GeV/c$^{2}$ are allowed \cite{lightmm,alvaro}.
Although it is not yet possible to set direct limits at this mass
scale, it is worthwhile to search in the accessible region at 
LEP energies.

Searches for classical point-like monopoles have been performed 
mainly at high-energy accelerators and in cosmic radiation
experiments. 
Monopole searches have predominantly used either ionization or
induction detection techniques. 

Induction experiments measure the monopole magnetic charge and are
independent of monopole mass and velocity. These experiments 
search for
the induction of a persistent current within a superconducting loop 
\cite{loop}. Searches for magnetic monopoles using this method have 
been performed at the p$\bar{\mathrm p}$ Tevatron collider assuming that 
produced MMs could stop, and be trapped and bound, in the matter 
surrounding the D0 and CDF collision regions \cite{trapped}. The same 
strategy has been used to search for magnetic monopoles produced in
e$^{+}$p collisions at HERA \cite{hera}.

Ionization experiments rely on the large magnetic charge of monopoles
to produce more ionization than an electrical charge travelling with the same
velocity.
For $g=g_{D}$ and velocities $\beta =\left(v/c\right) \geq 10^{-2}$ a
magnetic monopole behaves, in terms of ionization energy loss
$(dE/dx)$, like an equivalent electric charge with $(ze)_{eq}=g_{D}
\beta$. The energy losses are thus very large 
\begin{equation}
    (dE/dx)_{g}=(g \beta /e)^{2}(dE/dx)_{e}
\label{Eq.dedx}
\end{equation}
and Dirac magnetic monopoles would be easily distinguished
from minimum ionizing electrically charged Standard Model (SM)
particles \cite{dedx,GG,lepOLD}.
Direct searches for magnetic monopoles using tracking devices
were performed at p$\bar{\mathrm p}$ and \epem\ colliders.
Experiments at the Tevatron collider established cross section limits of about
$2\times 10^{-34}$~cm$^{2}$ for MMs with $m_{M} < 850$~GeV/c$^{2}$
\cite{Bertani}, while searches at LEP have excluded masses up to
45~GeV/c$^{2}$ \cite{LEP}.

Indirect searches for classical monopoles have relied on the effects
of virtual monopole/anti-monopole loops added to QED processes
in p$\bar{\mathrm p}$ and \epem\ collisions \cite{abbott, acciarri}. 
Since the Standard Model \Zz boson could couple to monopoles, 
assuming that the coupling between the \Zz\ and a MM pair is
larger than
for any lepton pair, the measurement of the \Zz\ decay width
provides an indirect limit on MM production
for $m_{M}<m_{Z}/2$ \cite{alvaro,lepOLD}.




This paper describes a direct search for MM
pairs produced in  $\epem \rightarrow M\bar{M}(\gamma)$ reactions.
The data were collected with the OPAL detector at the LEP accelerator 
at CERN.
This search was primarily
based on the $dE/dx$ measurements in the tracking chambers.
OPAL has a well established analysis to search for stable, 
long-lived, massive particles using the $\dedx$ signatures of 
individual charged particle tracks  \cite{benelli}.
This analysis technique could not be used here because MMs are too 
heavily ionizing, resulting in charge saturation in the central jet chamber.
Therefore, a new analysis
method was developed based on hit information rather than 
reconstructed tracks.
The analysis was sensitive to MMs with masses from 45~GeV/c$^{2}$ up
to the kinematic limit (about 103~GeV/c$^{2}$).

\section{The OPAL Detector}

A description of the OPAL detector and its jet chamber can be found
in reference \cite{ref:detector}.
Only a brief overview is given here.

The OPAL detector operated at LEP between 1989 and 2000 and is now
dismantled.
The central detector comprised a system of tracking chambers,
providing track reconstruction over 96\% of the full solid
angle\footnote
   {The OPAL right-handed coordinate system is defined such that the
$z$-axis is 
    in the
    direction of the electron beam, the $x$-axis points toward the
centre of the 
    LEP ring, and $\theta$ and $\phi$ are the polar and azimuthal
angles, defined 
    relative to the $+z$- and $+x$-axes, respectively. The radial
coordinate is 
    denoted by $r$.}
inside a 0.435~T uniform magnetic field parallel to the beam axis. 
It consisted of a two-layer
silicon microstrip vertex detector, a high-precision vertex drift
chamber with axial and stereo wires,
a large-volume jet chamber and a set of $z$-chambers measuring 
the track coordinates along the beam direction. 

The jet chamber (CJ) \cite{CJ} is the most important detector for this analysis.
The chamber, with a diameter of about 2m and a length of about 4m, was divided
into 24 azimuthal sectors, each equipped with 159  
sense wires.  Up to 159 position and $\dedx$
measurements per track were thus possible. 

The CJ also provided the hardware trigger for monopole candidates.
This trigger identified events with highly ionizing particles. 
Of the 159 sense wires of a sector, 36 wires were combined to define 
three groups 
with 12 wires each. One group was at an inner region, close to the 
\epem collision axis. The other two groups were at central and outer 
regions. For each wire, 
hits from highly ionizing tracks were identified as those yielding 
an integrated signal above a threshold of 1250 counts in the Flash 
Analogues to Digital Converters (FADC). 
For comparison, a minimum ionizing particle yields about 200 FADC 
counts.
Values slightly above 1000 FADC counts 
are typical for protons with a momentum of a few hundred MeV. 
If, within a group, more than 10 wires detected a high dE/dx hit, a decision bit was set. 
If this bit was set by all groups of a sector, the monopole trigger was fired.
Using raw hit information of randomly triggered events, the monopole trigger 
was determined to have an efficiency greater than 99\%.



A lead-glass electromagnetic 
calorimeter located outside the magnet coil
covered the full azimuthal range with good hermeticity
in the polar angle range of $|\cos \theta |<0.984$.
The magnet return yoke was instrumented for hadron calorimetry
covering the region $|\cos \theta |<0.99$ and was surrounded by
four layers of muon chambers.
Electromagnetic calorimeters close to the beam axis 
completed the geometrical acceptance down to 24 mrad
on each side of the interaction point.
These small-angle calorimeters were also used to measure the
integrated luminosity by counting Bhabha events \cite{lumi-paper}.

In order to trigger on the signal described in the
introduction, only data collected when the monopole trigger was active 
were used. 
The data-set analysed here was recorded during the LEP2 phase
with an average centre-of-mass (c.m.) energy of 206.3~GeV, and corresponded to a
total integrated luminosity of 62.7~pb$^{-1}$.

\section{Monte Carlo Simulation}

The signal reaction $\mathrm{e}^{+}\mathrm{e}^{-}\rightarrow
M\bar{M}$ 
was simulated at $\sqrt{s_{MC}}$~=~208~GeV for 
monopole masses ($m_{M}$) of
45, 50, 55, 60, 65, 70, 75, 80, 85, 90, 95, 100, 101, 102, 103 and
104~GeV/c$^{2}$ with Monte Carlo (MC) event samples.
Each sample contained 1000 events.  
Small differences in the centre-of-mass energies between the 
OPAL data analysed ($\sqrt{s_{min}}$=~203.6~GeV, 
$\sqrt{s_{max}}$=~207.0~GeV, for an average
$\sqrt{s_{data}}$=~206.3~GeV) and the signal MC samples
($\sqrt{s_{MC}}$) have a negligible effect on the analysis. MM masses
were scaled to the c.m. energy with the equation:
\begin{equation}
m_{scaled} = m_{M}\sqrt{\frac{s_{data}}{s_{MC}}}
\end{equation}
This scaling is valid 
since $\dedx$ (hence detection efficiency) is a linear function of
mass.  

The very large value of the magnetic charge makes it impossible to 
use perturbative theory to calculate the MM production process.
MMs were assumed to be spin 1/2 particles, produced from the \epem\,
initial state
via annihilation into a virtual photon, which yields a
monopole-antimonopole
pair with a uniform azimuthal distribution and with the typical
fermion polar angle distribution $\propto (1+ \cos^{2}\theta)$:
\begin{equation}
\mathrm{e}^{+}\mathrm{e}^{-}\rightarrow \gamma^{*} \rightarrow
M\bar{M} 
\end{equation}
%
%
Since magnetic charge cannot be simulated directly, MMs were 
simulated as heavy electrically charged fermions
with an effective charge of $(ze)_{eq}=g_{D} \beta$ (assuming $n=1$).
The specific ionization energy loss was computed according to Eq.~\ref{Eq.dedx}. 

A magnetic monopole interacts with a magnetic field analogously to
how an electron interacts with an electric field. The Lorentz force
for a magnetic monopole carrying magnetic charge $g$ is: 
\begin{equation}
\vec{F}=g\left(\vec{B}-\vec{v}\times\vec{E}\right)
\end{equation}

The GEANT3 \cite{geant} based OPAL detector simulation program 
\cite{gopal}
was used to simulate the behavior of the MMs in the OPAL detector.
The routines to transport the particles through the magnetic field were modified 
such that over a given step the change in the momentum $d\vec{p}/dt$ of the monopole 
was obtained by solving analytically the differential equation: 
\begin{equation}
\frac{d\vec{p}}{dt} = g\vec{B}
\end{equation}

The solution describes the motion of a magnetic monopole in a uniform
magnetic field. The trajectory is a parabola, 
accelerating in the direction of the magnetic field. In
the plane perpendicular to the magnetic field the motion is along a
straight line, in sharp contrast to electrically charged particles,
which curve in this plane. 

We studied the effects of multiple scattering of the monopoles and the 
modelling of the electric field between the anode, cathode, and 
potential wires in CJ and found them to be negligible.

 
A software emulation of the monopole trigger was used to study its efficiency. 
For the simulated monopole events, the trigger efficiency was found 
to be essentially 100\%.

The background was estimated using MC simulations of Standard Model
processes, generated at $\sqrt{s}$=206~GeV. 
Two-fermion events ($Z^{0*}/\gamma^{*}\rightarrow
f\overline{f}(\gamma)$ with $f$ = $e, \mu,\tau, q$) were simulated
with KK2f \cite{twofermions}. For the two-photon background, the
PYTHIA \cite{pytia} and PHOJET \cite{phojet} Monte Carlo generators
were used for $\epem q\overline{q}$ final states and the Vermaseren
\cite{vermaseren} and BDK \cite{bdk} generators for all $\epem
l^{+}l^{-}$ final states. Four-fermion final states were simulated
with grc4f \cite{grc4f}, which takes into account interference
between all diagrams.

All generated signal and background events were processed through the
full simulation of the OPAL detector. The same event
analysis chain was applied to the simulated events and to the data. 

\section{Data Analysis}
\label{sect.4}

\begin{table}
\begin{center}
{\small
\begin{tabular}{|c|l|l|} \hline
Cut & Description & cut value \\\hline
Preselection & Total charge per hit (CJ):        & $\ge$  1000~FADC \\
             & Number of Tracks plus Clusters:   & $\le$ 18 \\\hline
    1        
             & The first hit wire:               & $\le$ 2 \\
             & Number of Tracks plus Clusters:   & $\le$ 4 \\\hline
    2        & Distance between the 2 sectors:   & $\ge$ 8 \\
    3        & Number of hits in overflow in $\majorsec$: & $\ge$ 10 \\
    4        & Z mean coordinate (CJ):           & $\le$  50~cm \\
    5        & Charge per hit in the $\majorsec$: & $\ge$  3700~FADC counts
\\
    6        & Charge per hit in the $\minorsec$: & $\ge$  3000~FADC counts
\\
             & Total charge per hit (CJ):        & $\ge$ 2500~FADC counts
\\\hline
\end{tabular}}
\parbox{0.9\textwidth}{\caption {\sl  
List of cuts applied to the data.
\label{tab:raw1}}}
\end{center}
\end{table}

Magnetic monopoles would distinguish themselves
by their anomalously high ionization energy loss in CJ and by
the different plane of curvature of the trajectory in the magnetic field,
compared to electrically charged particles.  

The large value of the specific energy loss ($\dedx$) of a MM in the
gas of
the tracking detectors would induce a saturation in most of the wire
hits.
With the signals from both ends of the wire saturated, it is not
possible to determine the $z$ position from charge sharing. In this 
case the $z$ position is set to zero by the reconstruction program.
In the MC, most MM events are seen to exhibit a mean z-coordinate near
zero, because of saturation effects.
Rather than trying to reconstruct the MM tracks in 3 dimensions,
events were examined for the characteristic MM pattern
of ionisation in the sectors of the OPAL Jet Chamber.


Pair-produced magnetic monopoles, $\epem\rightarrow M\bar{M}(\gamma)$,
would be expected to be produced back to back with a
characteristic pattern of hits 
in the jet chamber.
This would have resulted in an azimuthal separation of about 12 sectors
between the two sectors with the highest energy deposits,
called $\majorsec$ and $\minorsec$, with little energy deposited
elsewhere in the detector. 


Based on these considerations, events were rejected if the overall
charge deposited on the sense wires 
normalised per hit was smaller than 1000 FADC counts, or if the total multiplicity
of tracks plus clusters in the detector was greater than 18. 
The level of the FADC counts were based on gains and calibrations. We
refer to these two cuts as the preselection, see
Table~\ref{tab:raw1}.

To reject some un-modelled events, further cuts were applied:
the number of reconstructed tracks plus clusters
had to be no more than 4 and the first wire hit in CJ had to be one
of the first two wires (cut 1 in Table~\ref{tab:raw1}).
Table~\ref{tab:raw1} summarizes the other selection criteria.
We required the $\majorsec$ and $\minorsec$ to have an azimuthal 
separation of at least
eight sectors (cut 2) and the number of hits in overflow in the
$\majorsec$ to be larger than or equal to 10 (cut 3). Since 
the typical MM signature would exhibit a mean z-coordinate 
near zero, the
average of the $z$ coordinate in CJ was required to be less than 50~cm (cut
4). The deposited charge per hit in $\majorsec$ and $\minorsec$
was required to be larger than 3700 FADC and 3000 FADC counts, respectively (cut
5 and cut 6) and the total charge per hit in all the
CJ sectors to be larger than 2500 FADC counts (cut 6).

The Standard Model background was dominated by Bhabha events and 
two-photon hadronic events, with a contribution from other two-photon
events.
The effect of the cuts on the samples at an average 
c.m. energy of $\sqrt{s}$=206.3~GeV is
shown in Table~\ref{tab:raw2}.
After applying cut 1, there was poor agreement between data
and MC (see Table~\ref{tab:raw2}). 
This was because the  
data still contained remaining un-modelled backgrounds from beam-gas interactions,
cosmic rays and
detector noise. This un-modelled background was much reduced by the
subsequent cuts since beam-gas interactions yield particles
which mainly travel along the beam pipe and do not have the
characteristic back-to-back
pattern, and detector noise does not
deposit large amounts of charge on the wires.
The remaining difference of 15-20$\%$ between the number of events in
the data 
and MC after cut 2 does not affect our results, as the signal is so separated 
from the the background that we can impose very hard cuts to remove 
all the background without affecting the detection efficiency.

\begin{center}
\begin{figure}[h!]
\epsfig{file=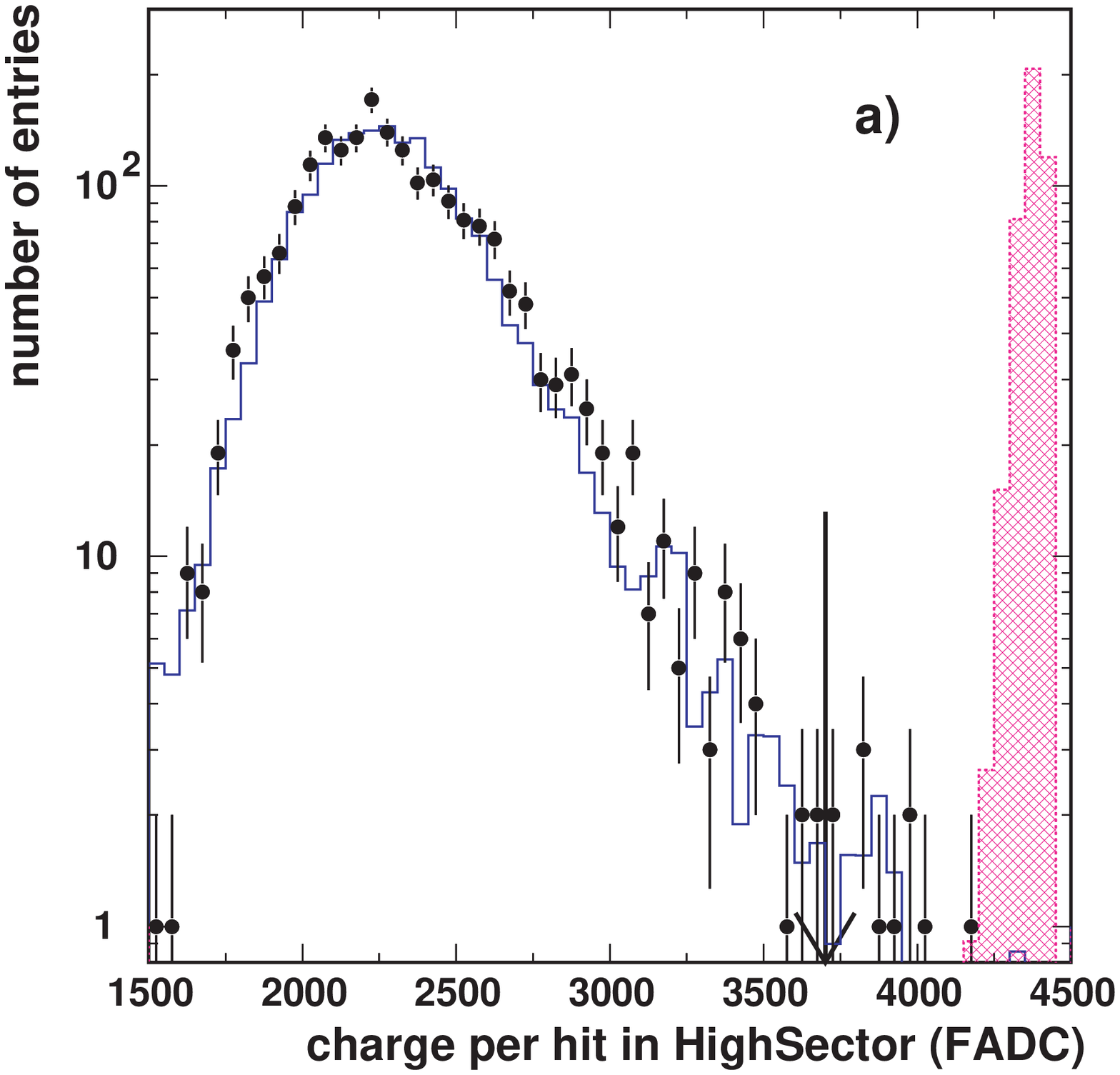,width=0.5\textwidth} 
\epsfig{file=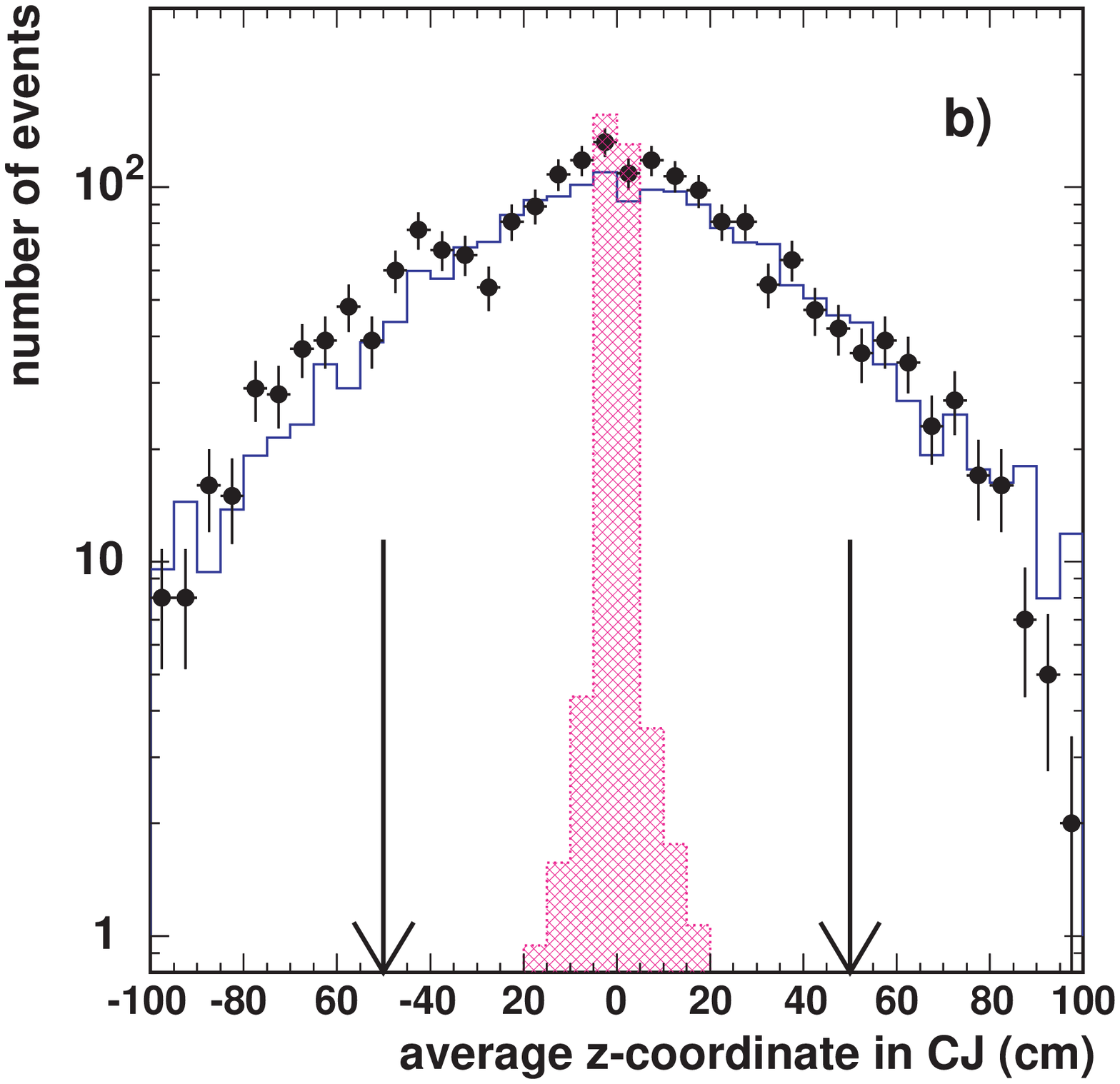,width=0.5\textwidth} 

\caption[]
        {Global event properties after applying cut 2.
The solid histogram is the generated Standard Model MC
background,
the filled histogram is the MC magnetic monopole signal and the 
points are the data. 
a) charge deposited in the $\majorsec$ in CJ (FADC); 
b) average z-coordinate of all hits in the jet chamber. 
Most MM events are seen to exhibit a mean z-coordinate near
zero, because of the saturation effects mentioned in
Sect.~\ref{sect.4}.
The arrows in each plot show the cuts applied in the
analysis.} 
\label{fig:variable}
\end{figure}
\end{center}

\begin{figure}[t]
\begin{center}
\epsfig{file=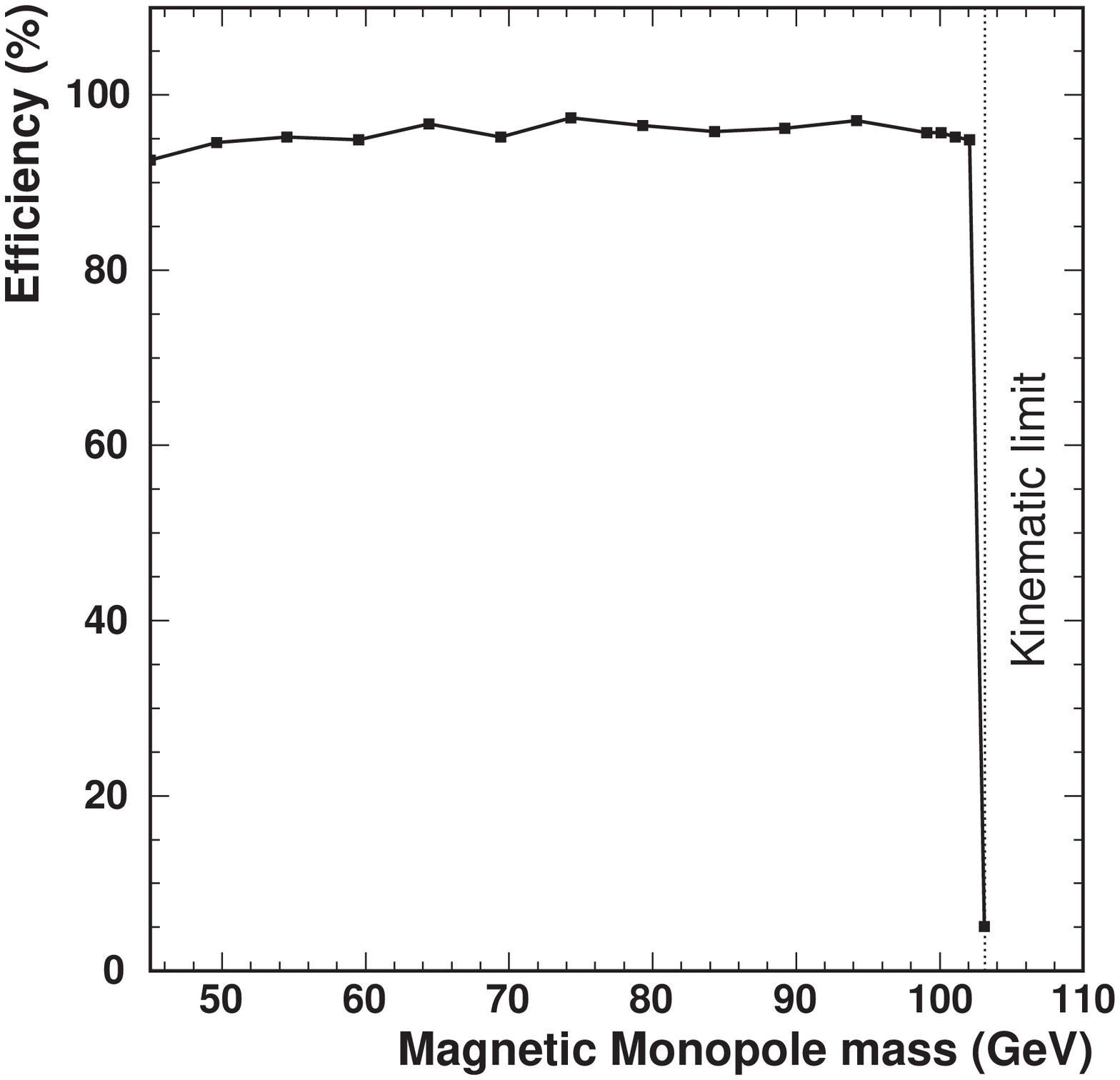,width=0.55\textwidth}
\caption[]
        {Monte Carlo estimate of the selection efficiency as a
	function of 
         the monopole mass at $\sqrt{s}$=206.3~GeV. The dotted line
	 is the kinematic limit.}
\label{fig:effy}
\end{center}
\begin{center}
\epsfig{file=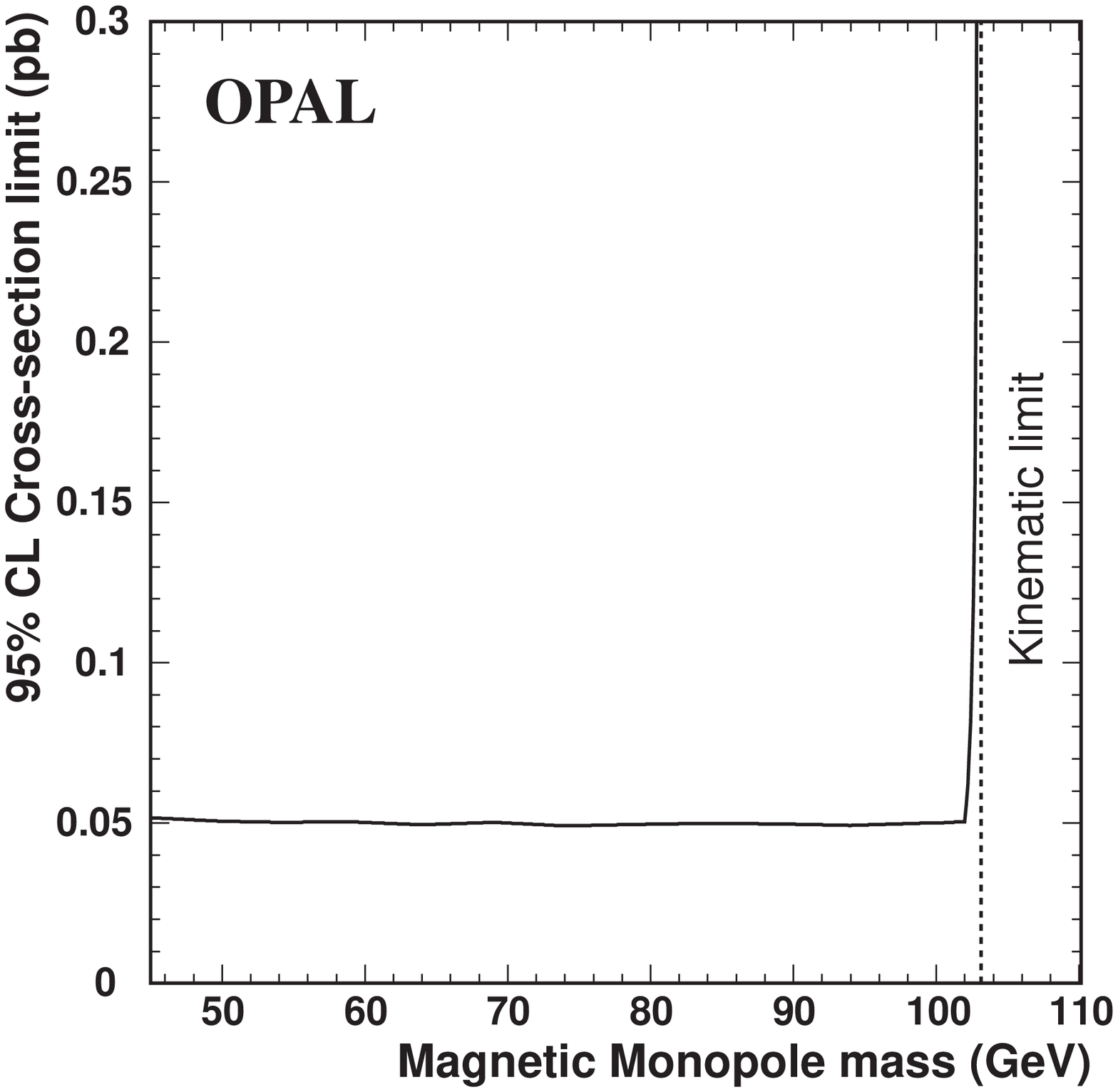,width=0.55\textwidth}
\caption[]
        {The model-independent 95$\%$ C.L. upper limit 
         on the pair-production 
         cross-section of magnetic monopoles in \epem\ collisions
	 at LEP2 at $\sqrt{s}$=206.3~GeV (plotted vs MM mass).}
\label{fig:limit}
\end{center}
\end{figure}

Fig.~\ref{fig:variable} shows the distribution of two of the main variables used by the analysis
after cut 2:
the charge per hit in the CJ sector $\majorsec$ and the average of
the z-coordinate. 
The total number of data events at this stage is 2928 and the total
number of
the MC Standard Model events is 2462 (Table~\ref{tab:raw2}). 
Since the magnetic monopole behavior would be very different from any
electrically charged SM particles, 
all the variables used by the analysis have a very well separated
distribution for the MM signal and SM MC backgrounds. 
For this reason it can be seen from Table~\ref{tab:raw2} that no MC
background event survived the analysis cuts. 
Moreover the overall detection efficiency is very high ($\geq 90\%$)
for almost all MM masses.
In Fig.~\ref{fig:effy} the detection efficiency for pair-produced
magnetic monopoles at $\sqrt{s}\cong$ 206~GeV is shown as a function
of m$_M$. 

 
\begin{table}
\begin{center}
{\small
\begin{tabular}{|ccc|c|c|c|c|c|c|c|c|c|c|} \hline
&&&
\multicolumn{9}{|c|}{Number of background events SM MC} & \\
\hline
cut & data & Total SM MC & bhabha & 2f & qq & $2\gamma$(e) &
$2\gamma(\mu)$ & $2\gamma(\tau)$ & $\nu\nu$ & 4f & $2\gamma$(q) &
sig. eff.($\%$)  \\ \hline
 1  &  44491 &  5707 &  4231 & 0.7 & 0.6 & 75.3 &  2.2 &   71.9 &
1.9&    57 &  1266 & 91 \\ 
 2  &   2928 &  2462 &  1927 & 0.1 & 0.3 &  6.0 &  0.1 &   27.5 &
0.3 &    14 &   487 & 91 \\ 
 3  &   2576 &  2194 &  1661 & 0.1 & 0.3 & 5.4 &  0.1 &    27.4 &
0.3 &  12.8 &   487 & 91 \\   
 4  &   1982 &  1597 &   1405& 0.0 & 0.0 &  0.4 &  0.0 &   6.9 &
0.0 &   6.4 &   177 & 91 \\
 5  &      2 &   1.2 &   0.5 & 0.0 & 0.0 &  0.0 &  0.0 &   0.1 &
0.0 &   0.0 &   0.6 & 91 \\   
 6  &      0 &   0.0 &   0.0 & 0.0 & 0.0 &  0.0 &  0.0 &   0.0 &
0.0 &   0.0 &   0.0 & 91 \\\hline

\end{tabular}}
\parbox{0.9\textwidth}{\caption {\sl
The number of data and Monte Carlo events remaining after the cuts
for analysed data-set collected 
at $\sqrt{s}$=206.3~GeV and for various 
MC SM background processes normalised to the integrated luminosity of
the data (62.7~pb$^{-1}$).
The last column gives the  efficiencies (in percent) for the magnetic
monopole MC signal 
simulated in the mass region between 45~GeV/c$^{2}$ and
103~GeV/c$^{2}$. 
\label{tab:raw2}}}
\end{center}
\end{table}

\section{Estimates of Systematic Uncertainties}

The distributions of the variables in the data and SM MC have similar shapes. 
The differences in the mean values are quite small.
The MC modelling of the $\dedx$ may introduce some systematic uncertainties. 
These
%
were evaluated by displacing the cut value on a given variable $x$
from the original position $x_{0}$ to a new position
$\overline{x}_{0}$, to reproduce on the simulated events the effect
of the cut on the real data. $\overline{x}_{0}$ is defined by:

\begin{equation}
        \overline{x}_{0}=\left(x_{0}-\left\langle
x\right\rangle_{data}
\right)\frac{\sigma_{bkg}}{\sigma_{data}}+\left\langle
x\right\rangle_{bkg}
\end{equation}
where $\left\langle x\right\rangle_{data}$, $\left\langle
x\right\rangle_{bkg}$, $\sigma_{data}$ and $\sigma_{bkg}$ are the
mean values and the standard deviations of the distributions of the
variable $x$ for the data and the simulated background. These
quantities were calculated from the $x$ distributions of the events
surviving the cuts on all the other variables used in the selection.
It was verified that using the distribution of $x$ at other stages of
the selection leads to negligible changes in the values of this
uncertainty.

The procedure was repeated for the main variables used in the event
selection (Table~\ref{tab:raw1}): the number of overflows in 
$\majorsec$, the Z mean coordinate in CJ and the charge per hit in 
$\majorsec$ and $\minorsec$. The difference between the reduced
efficiency, due to the displacement of the cut, and  that obtained
with the nominal selection was taken as the systematic uncertainty
due to the modelling of the variable under consideration. The
relative systematic uncertainties in the signal efficiency associated
with the various quantities are reported in Table~\ref{tab:sys}. The
range comes from different values obtained for the different MM
masses.

At a given centre-of-mass energy the different systematic
uncertainties were assumed to be independent, so that the total
systematic uncertainty was calculated as the quadratic sum of the
individual uncertainties. The global systematic uncertainty ranges
between 0.4\% and 5.2\% (Table~\ref{tab:sys}).

The MC statistical uncertainty, due to the limited number of signal
events generated, has been computed using a binomial formula and is
reported in Table~\ref{tab:sys}.

\begin{table}
\begin{center}
\begin{tabular}{|c|c|}\hline
Quantity & Systematic uncertainty (\%) \\\hline
Number of overflows in $\majorsec$ & 0.0 - 0.2 \\
Z mean coordinate (CJ)               & 0.2 - 0.4 \\
Charge per hit in $\majorsec$ & 0.3 - 4.7 \\
Charge per hit in $\minorsec$ & 0.2 - 2.2 \\\hline\hline
Global systematic uncertainty        & 0.4 - 5.2 \\\hline
Signal MC statistics                 & 0.6 - 0.8 \\\hline\hline
Total                                & 0.7 - 5.3 \\\hline
\end{tabular}
\parbox{0.9\textwidth}{\caption {\sl Summary of systematic 
uncertainties for the
signal efficiency of the various quantities used in the analysis. The
range of results corresponds to the values obtained for the different MM
masses.
\label{tab:sys}}}
\end{center}
\end{table}

\section{Results and Conclusions}

No magnetic monopole signal was found in this search. 
In Figure~\ref{fig:limit} the $95\%$ CL upper limit on the production
cross-section
at an average c.m. energy of $\sqrt{s}=206.3$~GeV is shown as a
function of the monopole mass. 
The average upper limit on the cross-section, computed using a 
frequentist approach, is 0.05~pb in the mass
range $45<m_M<102$~GeV/c$^{2}$. This limit is essentially independent 
of the mass in this range.

The computation of the cross-section is non-trivial. Nevertheless we 
expect the cross-section to be large.
The cross-section for the pair production of Dirac Magnetic Monopoles
computed assuming a naive tree-level coupling through an s-channel
virtual photon, according to the effective charge $(ze)_{eq}= g_{D}
\beta$, is around 5 orders of magnitude larger than the upper limit
obtained in this experiment \cite{LEP}. In this model we can thus
exclude classical MMs in the mass range 45-102 GeV/c$^{2}$.
This is a new excluded mass range for Dirac magnetic monopole
searches in \epem\ interactions.



\section{Acknowledgements}
\par
We particularly wish to thank the SL Division for the efficient operation
of the LEP accelerator at all energies
 and for their close cooperation with
our experimental group.  In addition to the support staff at our own
institutions we are pleased to acknowledge the  \\
Department of Energy, USA, \\
National Science Foundation, USA, \\
Particle Physics and Astronomy Research Council, UK, \\
Natural Sciences and Engineering Research Council, Canada, \\
Israel Science Foundation, administered by the Israel
Academy of Science and Humanities, \\
Benoziyo Center for High Energy Physics,\\
Japanese Ministry of Education, Culture, Sports, Science and
Technology (MEXT) and a grant under the MEXT International
Science Research Program,\\
Japanese Society for the Promotion of Science (JSPS),\\
German Israeli Bi-national Science Foundation (GIF), \\
Bundesministerium f\"ur Bildung und Forschung, Germany, \\
National Research Council of Canada, \\
Hungarian Foundation for Scientific Research, OTKA T-038240, 
and T-042864,\\
The NWO/NATO Fund for Scientific Research, the Netherlands.\\





\vspace{2cm}

\newpage

\end{document}